\documentclass{article}
\def\Ref#1{(\ref{#1})}
\usepackage{amsmath}
\usepackage{amssymb}
\usepackage{cite}

\begin{document}
\begin{titlepage}
\noindent{\large\textbf{phase transitions in systems possessing
Shock solutions}}

\vskip 2 cm
\begin{center}{Maryam~Arabsalmani{\footnote
{m\_arabsalmani@alzahra.ac.ir}} \& Amir~Aghamohammadi{\footnote
{mohamadi@alzahra.ac.ir}}  } \vskip 5 mm

\textit{     Department of Physics, Alzahra University,
             Tehran 19938-91167, Iran. }

\end{center}

\begin{abstract}
\noindent Recently it is shown that there are three families of
stochastic one-dimensional non-equilibrium lattice models for
which the single-shock measures form an invariant subspace of the
states of these models. Here, both the stationary states and
dynamics of single-shocks on a one-dimensional lattice are
studied. This is done for both an infinite lattice and  a finite
lattice with boundaries. It is seen that these models possess both
static and dynamical phase transitions. The static phase
transition is the well known low-high density phase transition For
the ASEP. The BCRW, and AKGP models also show the same phase
transition. Double-shocks on a one-dimensional lattice are also
investigated. It is shown that at the stationary state the
contribution of double-shocks with higher width becomes small, and
the main contribution comes from thin double-shocks.
\end{abstract}
\begin{center} {\textbf{PACS numbers:}}   02.50.Ga

\end{center}
\end{titlepage}
\section{Introduction}
Reaction-diffusion systems, is a well-studied area. People have
studied reaction-diffusion systems, using analytical techniques,
approximation methods, and simulations. A large fraction of exact
results  belong to low-dimensional (specially one-dimensional)
systems, where solving low-dimensional systems should in principle
be easier. Despite their simplicity, these systems exhibit a rich
and rather non-trivial dynamical and stationary behavior. Studies
on the models  far from equilibrium have shown that there is a
remarkably rich variety of critical phenomena\cite{ScR}.

Shocks in one-dimensional reaction-diffusion models have been
absorbed many interest recently
\cite{PS0,KJS,BS,DLS,PS,J1,PS2,JGM,J2,JM,RS}. There are some exact
results about shocks in one-dimensional reaction-diffusion models
together with simulations, numeric results \cite{J1} and also mean
field results \cite{PS0}. Formation of localized shocks in
one-dimensional driven diffusive systems with spacially
homogeneous creation and annihilation of particles has been
studied in \cite{PRWKGS}. Recently, in \cite{KJS},  the families
of models with travelling wave solutions on a finite lattice have
been presented. These models are the Asymmetric Simple Exclusion
Process (ASEP), the Branching- Coalescing Random Walk (BCRW) and
the Asymmetric Kawasaki-Glauber process (AKGP). In all of these
cases the time evolution of the shock measure is equivalent to
that of a random walker on a lattice with $L$ sites with
homogeneous hopping rates in the bulk and special reflection rates
at the boundary.

Shocks have been studied at both the macroscopic and the
microscopic levels and there are some efforts on addressing the
question that how these macroscopic shocks originate from the
microscopic dynamics \cite{PS2}. Hydrodynamic limits are also
investigated.

Among the important aspects of reaction-diffusion systems, is the
phase structure of the system. The static phase structure concerns
with the time-independent profiles of the system, while the
dynamical phase structure concerns with the evolution of the
system, specially its relaxation behavior. In
\cite{MA1,AM2,MAM,MA2}, the phase structure of some classes of
single- or multiple-species reaction-diffusion systems have been
investigated. These investigations were based on the one-point
functions of the systems.

Here we study both stationary and also dynamics of the
single-shocks on a one-dimensional lattice. This is done for both
an infinite lattice and a finite lattice with boundaries. In the
stationary state, the system can be found in the low-density or in
the high-density phase. This phase transition is a well known
first order phase transition in familiar ASEP, and also its
extensions \cite{phase}. The BCRW, and AKGP models show the same
phase transitions. We also investigate the dynamical phase
transitions of the models. It is seen that ASEP has no dynamical
phase transition, but both of the models  BCRW, and AKGP have
three phases, and system may show dynamical phase transitions.
Double shocks on a one-dimensional lattice have been also
investigated, and its the stationary behavior have been studied.
It is shown that, in the thermodynamic limit, contribution of
double shocks with higher width become vanishingly small.

\section{Fixing the notations}
Consider a one-dimensional lattice, each point of which is either
empty or contains one particle. Let the lattice have $L$ sites. An
empty state  is denoted by $|0\rangle $ and an occupied state is
denoted by $|1\rangle $.
\begin{equation}
|0\rangle : = \begin{pmatrix}1\cr 0\end{pmatrix}, \qquad |1\rangle
: = \begin{pmatrix}0\cr 1\end{pmatrix}.
\end{equation}
If the probability that the site $i$ is occupied is  $\rho_i$
then the state of that  is represented by
$\begin{pmatrix}1-\rho_i\cr \rho_i\end{pmatrix}$. The observables
of a reaction-diffusion system are the operators $N_i^\alpha$,
where $i$ with $1\leq i\leq L$ denotes the site number, and
$\alpha=0,1$ denotes the hole or the particle: $N_i^0$ is the
hole (vacancy) number operator at site $i$, and $N_i^1$ is the
particle number operator at site $i$. One has obviously the
constraint
\begin{equation}\label{1}
s_\alpha N^\alpha_i=1,
\end{equation}
where ${\langle s| }$ is a covector the components of which
($s_\alpha$'s) are all equal to one. The constraint \Ref{1},
simply says that each site is either occupied by one particle or
empty. A representation for these observables is
\begin{equation}\label{2}
N_i^\alpha:=\underbrace{1\otimes\cdots\otimes 1}_{i-1}\otimes
N^\alpha\otimes\underbrace{1\otimes\cdots\otimes 1}_{L-i},
\end{equation}
where $N^\alpha$ is a diagonal $2\times 2$ matrix the only nonzero
element of which is the $\alpha$'th diagonal element, and the
operators 1 in the above expression are also $2\times 2$ matrices.
The state of the system is characterized by a vector
\begin{equation}\label{4}
{|\mathbb P\rangle }\in\underbrace{{\mathbb
V}\otimes\cdots\otimes{\mathbb V}}_{L},
\end{equation}
where ${\mathbb V}$ is a $2$-dimensional vector space. All the
elements of the vector ${|\mathbb P\rangle }$ are nonnegative, and
\begin{equation}\label{5}
{\langle\mathbb S| }{\mathbb P\rangle }=1.
\end{equation}
Here ${\langle\mathbb S| }$ is the tensor-product of $L$ covectors
${\langle s| }$. The evolution of the state of the system is given
by
\begin{equation}\label{7}
\dot{|\mathbb P\rangle }={\mathcal H}\;{|\mathbb P\rangle },
\end{equation}
where the Hamiltonian ${\mathcal H}$ is stochastic, by which it is
meant that its nondiagonal elements are nonnegative and
\begin{equation}\label{8}
{\langle\mathbb S| }\; {\mathcal H}=0.
\end{equation}
Two conventions are used to write the master equation,
\begin{eqnarray}\label{07}
a)&\quad \dot{|\mathbb P\rangle }=-{\mathcal H'}\;{|\mathbb
P\rangle }\quad \Rightarrow ,\quad {|\mathbb P\rangle
}(t)=\exp\left(-t{\mathcal H'}\right)\;{|\mathbb P\rangle
}(0)\cr&\cr
 b)&\quad \dot{|\mathbb P\rangle }={\mathcal
H}\;{|\mathbb P\rangle }\quad \Rightarrow ,\quad {|\mathbb
P\rangle }(t)=\exp\left(t{\mathcal H}\right)\;{|\mathbb P\rangle
}(0)\nonumber
\end{eqnarray}
These two conventions are related to each other simply through
${\mathcal H'}=-{\mathcal H}$. In the $a)$ convention nondiagonal
elements of ${\mathcal H}'$ are negative of reaction rates, hence
nonpositive, and its diagonal elements are nonnegative. In the
$b)$ convention nondiagonal elements of ${\mathcal H}$ are
reaction rates, hence nonnegative, and  its diagonal elements are
nonpositive. In the $a)$ convention the real parts of the
eigenvalues of ${\mathcal H}'$ are nonnegative, and the eigenvalue
with minimum nonzero real part corresponds to the relaxation time.
In the $b)$ convention the real parts of the eigenvalues of H are
nonpositive, and the eigenvalue with maximum nonzero real part
corresponds to the relaxation time, and finally the state vector
${|\mathbb P\rangle }$ is the same in two conventions, and in both
conventions the elements of ${|\mathbb P\rangle }$ are
nonnegative. Through out this paper we use the $b)$ convention.

The interaction is nearest-neighbor, if the Hamiltonian is of the
form
\begin{equation}\label{9}
{\mathcal H}=\sum_{i=1}^{L-1}H_{i,i+1},
\end{equation}
where
\begin{equation}\label{10}
H_{i,i+1}:=\underbrace{1\otimes\cdots\otimes 1}_{i-1}\otimes H
\otimes\underbrace{1\otimes\cdots\otimes 1}_{L-1-i}.
\end{equation}
(It has been assumed that the sites of the system are identical,
that is, the system is translation-invariant. Otherwise $H$ in the
right-hand side of \Ref{10} would depend on $i$.) The two-site
Hamiltonian $H$ is stochastic, that is, its non-diagonal elements
are nonnegative, and the sum of the elements of each of its
columns vanishes:
\begin{equation}\label{11}
({\langle s| }\otimes{\langle s| })H=0.
\end{equation}
Here $H$ is a $4\times 4$ matrix (as the system under
consideration has two possible states in each site and the
interactions are nearest neighbor). The non-diagonal elements of
$H$ are nonnegative and equal to the interaction rates; that is,
the element $H^\alpha_\beta$ with $\alpha\ne\beta$ is equal to the
rate of change of the state $\beta$ to the state $\alpha$.
$\alpha$ and $\beta$, each represent the state of two adjacent
sites. For example if $\alpha=01$ and $\beta=10$, then
$H^\alpha_\beta$ is the rate of particle diffusion to the right.

The evolution equation of one-point function $\langle n_i\rangle$
($k$-point functions) depends on both one- and two-point functions
($k-1$-, $k$- and $k+1$-point functions). Generally this set of
evolution equations cannot be solved exactly. If one can obtain
the state of the system ${|\mathbb P\rangle }$ exactly, there is
no need to solve this set of evolution equations. In \cite{GS}, a
ten-parameter family of reaction-diffusion processes was
introduced for which the evolution equation of $k$-point
functions contains only $k$- or less- point functions. We call
such systems autonomous. The average particle-number in each site
has been obtained exactly for these models. In \cite{AAMS,SAK},
this has been generalized to multi-species systems and
more-than-two-site interactions.

Although generally, one cannot obtain the state of the system
${|\mathbb P\rangle }$ exactly, for a special choice of initial
states and of course with some constraints on reaction rates, one
may obtain the state of the system ${|\mathbb P\rangle }$. There
are three families of stochastic one-dimensional non-equilibrium
lattice models for which the single-shock measures are an
invariant subspace of the states of these models \cite{KJS}. If
the initial state of these models is a linear superposition of
shock measures then one can obtain the state of the system
${|\mathbb P\rangle }$ exactly. These models are the Asymmetric
Simple Exclusion Process (ASEP), the Branching- Coalescing Random
Walk (BCRW) and the Asymmetric Kawasaki-Glauber process (AKGP).

Let's  consider a one-dimensional lattice with $L$ sites. The
interaction is nearest-neighbor, if the Hamiltonian is of the form
\begin{equation}\label{12}
{\mathcal H}=b_1 \otimes {\mathbf  1}^{\otimes (L-1)}+
\left[\sum_{i=1}^{L-1}H_{i,i+1}\right] + {\bf 1}^{\otimes
(L-1)}\otimes b_L,
\end{equation}
where $H_{i,i+1}$ denotes interaction in the bulk and $b_1$ and
$b_L$ denote the interactions at the boundaries,
\begin{equation}\label{13}
    b_1:=\left(\begin{array}{cc}
        -\alpha &\ \gamma \\
      \ \alpha & -\gamma
      \end{array}\right)\qquad
       b_L:=\left(\begin{array}{cc}
        -\delta &\ \beta \\
     \ \delta & -\beta
      \end{array}\right).
\end{equation}
$\alpha$ and $\gamma$ ($\delta$ and $\beta$) are the rates of
injection and extraction at the first site (at the $L$'th site).
Each site may be occupied or vacant. We represent any
configuration of the system by the vector ${|E_a\rangle }$. So
the system is spanned by $2^L$ vectors,  ${|E_a\rangle }$
($a=1,2,\cdots 2^L$), and any physical state is a linear
combination of these vectors
\begin{equation}\label{14-0}
{|\mathbb P\rangle }= \sum_{a=1}^{2^L}{\mathcal P}_a {|E_a\rangle
}, \qquad {\rm where}\qquad \sum_{a=1}^{2^L}{\mathcal P}_a=1.
\end{equation}
${\mathcal P}_a$'s are nonnegative real numbers. ${\mathcal P}_a$
is the probability of finding the system in the configuration $a$.

It is said that  the state of the system is a single-shock at the
site $k$ if there is a jump in the density at the site $k$ and the
state of the system is represented by a tensor product of the
states at each site as
\begin{equation}
  |e_k\rangle= u^{\otimes k}\otimes v^{\otimes (L-k)},
\end{equation}
where
\begin{equation}
 u:=\begin{pmatrix}1-\rho_1\cr \rho_1\end{pmatrix} \quad v:=\begin{pmatrix}1-\rho_2\cr
 \rho_2\end{pmatrix}.
\end{equation}
It is seen that
\begin{equation}\label{5-1}
{\langle\mathbb S| }{e_k\rangle }=1.
\end{equation}
$|e_k\rangle$ represents a state  for which the occupation
probability for the first $k$ sites is $\rho_1$, and  the
occupation probability for the next $L-k$ sites is $\rho_2$. The
set $|e_k\rangle, k=0,1,\cdots L$ is not a complete set, but
linearly independent.

There are three families of stochastic one-dimensional
non-equilibrium lattice models, (ASEP,BCRW,AKGP), for which if the
initial state of these models is a linear superposition of shock
states, at the later times the state of the system ${|\mathbb
P\rangle }$ remains a linear combination of shock state. For
these models
\begin{equation}\label{17}
    {\mathcal H} |e_k \rangle= d_1 |e_{k-1} \rangle+d_2 |e_{k+1} \rangle-(d_1+d_2) e_{k}.
\end{equation}
where $d_i$'s are some parameters depending on the reaction rates
in the bulk, $\rho_1$ and $\rho_2$. So the span of $|e_k
\rangle$'s is an invariant subspace of ${\mathcal H}$, the
Hamiltonian of the above mentioned models. It should be noted that
the number of $|e_k \rangle$'s are $L+1$, and any physical state
is not necessarily expressible in terms of $|e_k \rangle$'s. For a
finite lattice with the injection and extraction at the boundaries
besides (\ref{17}), there are two other relations.
\begin{eqnarray}\label{17-2}
 {\mathcal H} |e_0\rangle&=& D_1 |e_1\rangle-D_1 |e_0\rangle\cr
 {\mathcal H} |e_L\rangle&=& D_2 |e_{L-1}\rangle-D_2 |e_L\rangle,
\end{eqnarray}
where $D_i$'s are two parameters generally depending on $\rho_1$
and $\rho_2$, and the reaction rates.

 Let's assume  that the
initial state of the system is a linear combination of shock state
\begin{equation}\label{18-0}
{|\mathbb P\rangle }(0)= \sum_{k=0}^{L}p_k(0) {|e_k\rangle }.
\end{equation}
$p_k$s, are not necessarily nonnegative, and so any of them  may
be greater than one. For such an initial state, the system
remains in the sub-space spanned by shock measures.
\begin{equation}\label{18-7}
{|\mathbb P\rangle }(t)= \sum_{k=0}^{L}p_k(t) {|e_k\rangle }.
\end{equation}
 Using
(\ref{5-1}), it is seen that
\begin{equation}\label{5-2}
\sum_{k=0}^{L}{p}_k(t)=1.
\end{equation}
But it should be noted that  these  are not probabilities. $p_k$
only expresses the contribution of a shock at the site $k$ in the
state of the system. Any shock state ${|e_k\rangle }$ can be
expanded in terms of ${|E_a\rangle }$,
\begin{equation}\label{18-1}
{|e_k\rangle }= \sum_{a=1}^{2^L}\Lambda_{ka} {|E_a\rangle },
\end{equation}
where the elements of ${\bf \Lambda}$ are nonnegative.
Substituting (\ref{18-1}) in (\ref{18-7}) and comparing with
(\ref{14-0}) gives
\begin{equation}\label{18-2}
  {\mathcal P}_a= \sum_{k=1}^L p_k \Lambda_{ka}.
\end{equation}
Here ${\mathcal P}_a$ is the probability to find the system in the
state $|E_a\rangle$, and so it is a nonnegative  number. The
condition of non-negativeness of probabilities (${\mathcal P}_a$s)
leads to constraints on $p_k$, see (\ref{18-2}).

 The three models are classified as following
\begin{itemize}
    \item 1. {\underline{\rm ASEP}}- The only non-vanishing rates in the bulk
    are the rates of diffusion to the right $\omega_{23}$ and diffusion to the
    left $\omega_{32}$. For a finite lattice, there may be injection and
    extraction rates at the boundaries. $\alpha$, and $\gamma$ are
    the injection and extraction rates at the left boundary,
    and $\delta$, and $\beta$ the injection and extraction rates at the right boundary.
    In this case the densities can take any value
    between $0$ and $1$ ($\rho_1\ne 0,1 \quad \rho_2\ne 0,1$). $d_1$, and $d_2$
    are
\begin{align}\label{19}
d_1&=
\frac{\rho_1(1-\rho_1)}{\rho_2-\rho_1}(\omega_{23}-\omega_{32})\nonumber\\
d_2&=
\frac{\rho_2(1-\rho_2)}{\rho_2-\rho_1}(\omega_{23}-\omega_{32}).
\end{align}
It should be noted that the densities $\rho_1$, and $\rho_2$ are
also related through
\begin{equation}\label{20}
 \frac{\rho_2(1-\rho_1)}{\rho_1(1-\rho_2)}=\frac{\omega_{23}}{\omega_{32}}.
\end{equation}
So
\begin{equation}\label{20-1}
 d_1= \frac{\rho_1}{\rho_2}\omega_{23},\qquad d_2=
 \frac{\rho_2}{\rho_1}\omega_{32}.
\end{equation}

    The rates of injection and extraction at the boundaries are also
    related to the densities $\rho_1$, and $\rho_2$.
\begin{align}\label{21}
\rho_1(1-\rho_1)(\omega_{23}-\omega_{32})&=\alpha(1-\rho_1)-\gamma
\rho_1,\cr
-\rho_2(1-\rho_2)(\omega_{23}-\omega_{32})&=\delta(1-\rho_2)-\beta
\rho_2.
\end{align}
The parameters $D_1$, and $D_2$, are obtained to be
\begin{align}\label{21-2}
D_1&= d_2+\frac{\alpha (1-\rho_2)-\gamma \rho_2}{\rho_1-\rho_2}=
d_2+\frac{\alpha}{\rho_1}-\omega_{23} ,\cr D_2&= d_1-\frac{\delta
(1-\rho_1)-\beta\rho_1}{\rho_1-\rho_2}=d_1+\frac{\delta}{\rho_2}-\omega_{32}.
\end{align}

    \item 2. {\underline{\rm BCRW}}- The non-vanishing rates are coalescence ($\omega_{34}$, and $\omega_{24}$), Branching
    ($\omega_{42}$, and $\omega_{43}$) and  diffusion to the left and right ($\omega_{32}$, and $\omega_{23}$).
    The injection rate at the right boundary $\delta$ should be zero. The density  $ \rho_1$ can take any value
    between $0$ and $1$, but $\rho_2$ should be zero. These
    parameters are related through
\begin{align}\label{21-3}
\frac{\omega_{23}}{\omega_{43}}&=\frac{1-\rho_1}{\rho_1},\cr
\frac{\omega_{23}}{\omega_{43}}&=\frac{\omega_{24}+\omega_{34}}{\omega_{42}+\omega_{43}}.
\end{align}
 The rates of injection and extraction at the boundaries are also
    related to the densities $\rho_1$, and the reaction rates,
\begin{equation}\label{21-3-2}
\rho_1\omega_{23}-\rho_1(1-\rho_1)\omega_{32}-\rho_1^2\omega_{34}=\alpha(1-\rho_1)-\gamma
\rho_1.
\end{equation}
The parameters $d_1$, and $d_2$ are
\begin{align}\label{21-4}
d_1&= (1-\rho_1)\omega_{32}+\rho_1\omega_{34},\cr
d_2&=\frac{\omega_{43}}{\rho_1}.
\end{align}
The parameters $D_1$, and $D_2$, are obtained to be
\begin{align}\label{21-5}
D_1&= \frac{\alpha}{\rho_1} ,\cr D_2&= d_1-d_2(1-\rho_1)+\beta.
\end{align}
    \item 3. {\underline {\rm AKGP}}- The non-vanishing rates are
    Death ($\omega_{12}$, and $\omega_{13}$) and Branching to the left and right
    ($\omega_{42}$, and $\omega_{43}$), and also diffusion to the left.
    $\rho_1$ should be equal to one, and $\rho_2$ should be zero.
    The extraction rate at the left boundary $\gamma$,  and the injection rate at the right  boundary $\delta$
    should be zero. The hoping parameters are  $d_1= \omega_{13}$,
    $d_2=\omega_{43}$, and finally
    $D_1=\alpha$, and $D_2=\beta$.

\end{itemize}
Interchanging $\rho_1 $ and $\rho_2$, is nothing but exchanging
left and right. Changing $\rho_i$ to $1-\rho_i$ is particle hole
exchange.

\section{single-shock}

\subsection{Shock on an infinite lattice}
Here, we want to consider the evolution of shock measures on an
infinite lattice. Here $|e_l\rangle$ stands for a state with a
shock at the site $l$. If the initial state is a linear
combination of shock measures, at later times the state of the
system should be also expressible in terms of shock measures.
\begin{equation}\label{inf1}
{|\mathbb P\rangle }(t)= \sum_{k=-\infty}^{\infty}p_k(t)
{|e_k\rangle }.
\end{equation}
Calculating $p_{k}$, one can obtain any correlation functions of
number operators.
\begin{equation}\label{inf1-1}
    \langle n_i\rangle ={\langle\mathbb S| } n_i {|\mathbb P\rangle }=
    \sum_{k=-\infty}^{\infty}p_k(t){\langle\mathbb S| } n_i {|e_k\rangle },
\end{equation}
where
\begin{equation}\label{inf1-3}
{\langle\mathbb S| } n_i {|e_k\rangle }=\begin{cases}\rho_1&     i\leq k \\
                                            \rho_2 & i>k
\end{cases}.
\end{equation}
So\begin{equation}\label{inf1-4}
    \langle n_i\rangle = \rho_2 +(\rho_1-\rho_2){\mathcal B}_i(t)
\end{equation}
where
\begin{equation}\label{inf1-5}
   {\mathcal B}_i(t)=\sum_{k=i}^\infty p_k(t).
\end{equation}
All other correlation functions of number operators can be
obtained in terms of ${\mathcal B}_i$s.

\begin{eqnarray}
 \langle n_i\rangle&=&\rho_2+(\rho_1-\rho_2){\mathcal B}_i(t)\cr
\langle n_in_j\rangle&=&\rho_2^2+(\rho_1-\rho_2)\{\rho_2{\mathcal
B}_i(t)+\rho_1{\mathcal B}_j(t)\}\cr \langle
n_in_jn_k\rangle&=&\rho_2^3+(\rho_1-\rho_2)\{\rho_2^2{\mathcal
B}_i(t)+\rho_2\rho_1{\mathcal B}_j(t)+\rho_1^2{\mathcal
B}_k(t)\}\cr &&\hskip-1.3cm\vdots
\end{eqnarray}
where $i<j<k<\cdots$.

 The evolution equation for the system is
\begin{equation}\label{inf2}
{\mathcal H}{|\mathbb P\rangle }=\frac {\rm d}{{\rm d}t}{|\mathbb
P\rangle }.
\end{equation}
Knowing the action of ${\mathcal H}$ on ${|e_k\rangle }$, one can
obtain the evolution equation for $p_k(t)$.
\begin{equation}\label{inf3}
 \dot p_k= d_1 p_{k+1}+d_2 p_{k-1}-(d_1+d_2) p_{k}.
\end{equation}
Here we have used the linear independence of ${|e_k\rangle }$'s.
Let's define the generating function
\begin{equation}\label{inf4}
 G(z,t):= \sum_{l=-\infty}^{\infty}P_l(t)z^l.
\end{equation}
Then the evolution equation for $G(z,t)$ is
\begin{equation}\label{inf5}
 \dot G= [d_1 z^{-1}+d_2 z-(d_1+d_2)]G,
\end{equation}
the solution for which is
\begin{equation}\label{inf6}
  G(z,t)=e^{[d_1 z^{-1}+d_2 z-(d_1+d_2)]t}G(z,0).
\end{equation}
$G(z,0)$ can be determined using (\ref{inf4}) and  contributions
of the shock measures  in the initial state. The coefficients for
the Laurent expansion of the generating function are $p_k(t)$s;
\begin{equation}\label{inf7}
p_k(t) =e^{-(d_1+d_2)t}\sum_{m=-\infty}^{\infty}\left(\frac{d_2}{
d_1}\right)^{(k-m)/2}I_{k-m}(2\sqrt{d_1d_2} t)p_m(0),
\end{equation}
The above result is first obtained in \cite{BS}. At large times,
\begin{equation}\label{inf8}
p_k (t)
\sim\left(\frac{d_2}{d_1}\right)^{k/2}\frac{e^{[-(d_1+d_2)+2\sqrt{d_1d_2}]t}}{
\sqrt{t}}.
\end{equation}
It is seen from the above equation that if $d_2<d_1$, the
contribution of the shocks at the rightmost sites tend rapidly to
their final value, and obviously for $d_2>d_1$ the contribution of
the  shocks at the leftmost sites arrive earlier to their final
values. This expression seems to be unbounded for  $k\to \pm
\infty$. For any fixed $t$, this is true. However, it simply means
that in order that this term represents the leading term for some
$k$, $t$ must be greater than some $T$, which does depend on $k$.

\subsection{Shocks on a lattice with the boundary}
Now let's first consider a one-dimensional lattice with $L$ sites.
There are injection and extraction at the boundaries. Because of
the boundary terms (\ref{17}) changes to
\begin{eqnarray}\label{boun1}
 {\mathcal H} |e_k \rangle&=& d_1 |e_{k-1} \rangle+d_2 |e_{k+1}\rangle-(d_1+d_2) |e_k\rangle\cr
 {\mathcal H} |e_0\rangle&=& D_1 |e_1\rangle-D_1 |e_0\rangle\cr
 {\mathcal H} |e_L\rangle&=& D_2 |e_{L-1}\rangle-D_2 |e_L\rangle
\end{eqnarray}
The  evolution equation for $p_k$s in the bulk are the same as
that of infinite lattice, but here one should take care of
boundary terms.
\begin{eqnarray}\label{boun2}
\dot p_k&=& d_1 p_{k+1}+d_2 p_{k-1}-(d_1+d_2) p_k, \qquad k\ne
0,1,L-1,L\cr
 \dot p_{L-1}&=& D_2 p_L+d_2 p_{L-2}-(d_1+d_2) p_{L-1},\cr\dot
p_1&=& d_1p_2+D_1 p_0-(d_1+d_2) p_1,\cr \dot p_{L}&=& d_2
p_{L-1}-D_2 p_L,\cr \dot p_0&=& d_1 p_1-D_1p_0.
\end{eqnarray}
Let's first consider the stationary case.  These set of equations
can be solved easily,
\begin{equation}\label{boun3}
    p_k= (\frac{d_2}{
    d_1})^{k-1}(\frac{d_1}{D_2})^{\delta_{k,L}}(\frac{d_2}{
    D_1})^{\delta_{k,0}}p_1
\end{equation}
For any finite $L$, all the $p_k$s can be obtained in term of
$p_1$, and $p_1$ is also obtained using the normalization
condition (\ref{5-2}). In the thermodynamic limit ($L\to \infty$),
it is seen that for $d_2>d_1$, finiteness of $p_k$ for large $k$,
leads to vanishingly small $p_1$. But as $d_1$ exceeds $d_2$,
$p_k$ for large $k$ becomes vanishingly small in the thermodynamic
limit. This is the static phase transition previously mentioned.
The static phase transition is controlled by the reaction rates in
the bulk and is independent of the rates at the boundaries. It is
a discontinuous change of the behavior of the derivative of the
stationary value of $p_k$ at the end points, with respect to the
reaction rates.  This phase transition is a well known first order
phase transition for the  familiar ASEP, and also its extensions
\cite{phase}. The BCRW, and AKGP models shows the same phase
transitions. In the stationary state, the system can be found in
the low-density or in  the high-density phase, or in the other
words the contribution of leftmost or rightmost shocks may be
negligible.

Now, return to the dynamics of the system. To proceed let's first
introduce a change of variable
\begin{align}\label{boun4}
&q_k:=p_k,\qquad k\ne 0,L\nonumber \\
&q_0:=(D_1/d_2)p_0,\nonumber \\
&q_L:=(D_2/d_1)p_L.
\end{align}
Using this change of variable (\ref{boun2}) recasts to
\begin{eqnarray}\label{boun5}
&&\dot q_k= d_1 q_{k+1}+d_2 q_{k-1}-(d_1+d_2) q_k, \qquad k\ne
0,L\cr &&\cr &&\dot q_{L}= \frac{d_2D_2}{ d_1} q_{L-1}-D_2 q_L,\cr
&&\cr &&\dot q_0= \frac{d_1D_1}{ d_2} q_1-D_1q_0.
\end{eqnarray}
To find the relaxation of the system towards its stationary state,
one should find the greatest nonzero eigenvalue of the operator
$h$, defined through $\dot q_k=: h_k^lq_l$. The eigenvalues and
eigenvectors of $h$ have been denoted by $E$ and ${\bf C}_E$,
respectively. Expanding the vector $q$ in terms of ${\bf C}_E$'s,
and regarding the completeness and linear independency of ${\bf
C}_E$'s,  one arrives at
\begin{eqnarray}\label{boun7}
&&E\ C_k= d_1 C_{k+1}+d_2 C_{k-1}-(d_1+d_2) C_k, \qquad k\ne
0,L\cr &&\cr &&E\ C_{L}= \frac{d_2D_2}{ d_1} C_{L-1}-D_2 C_L,\cr
&&\cr &&E\ C_0= \frac{d_1D_1}{d_2} C_1-D_1C_0.
\end{eqnarray}
The solution to the above equations is
\begin{equation}\label{boun8}
C_k =a z_1^k+b z_2^k,
\end{equation}
where $z_i$'s satisfy
\begin{equation}\label{boun9}
E=-(d_1+d_2)+d_1z_i+\frac{d_2}{{z_i}}.
\end{equation}
Then  $z_1 z_2=d_2/d_1$. The second and third equations of
\Ref{boun7} take the form
\begin{eqnarray}\label{boun10}
(E+D_2)(a z_1^L+b z_2^L)-\frac{d_2D_2}{ d_1}(a z_1^{L-1}+b
{z_2}^{L-1})=0,\cr (E+D_1)(a+b)-\frac{d_1D_1}{ d_2}(a z_1+b
z_2)=0.
\end{eqnarray}
or
\begin{eqnarray}\label{boun11}
&&\left[(E+D_2)z_1-\frac{d_2D_2}{ d_1}\right]a+
\left[(E+D_2)z_2-\frac{d_2D_2}{
d_1}\right]\left(\frac{z_2}{z_1}\right)^{L-1}b=0,\cr
&&\left[E+D_1-\frac{d_1D_1}{
d_2}z_1\right]a+\left[E+D_1-\frac{d_1D_1}{ d_2}z_2\right]b=0.
\end{eqnarray}
To have nonzero solutions for  $q_k$'s, these equations should
have nontrivial solutions for $a$, and $b$, which means that the
determinant of the coefficients should be zero,
\begin{eqnarray}\label{boun12}
&&\left[E+D_1-\frac{d_1D_1}{
d_2}z_1\right]\left[(E+D_2)z_2-\frac{d_2D_2}{
d_1}\right]\left(\frac{z_2}{ z_1}\right)^{L-1}\cr &&
=\left[E+D_1-\frac{d_1D_1}{
d_2}z_2\right]\left[(E+D_2)z_1-\frac{d_2D_2}{d_1}\right].
\end{eqnarray}
Performing the change of variable $z_i=:\sqrt{d_2/d_1}\ x_i$,
leads to
\begin{equation}\label{boun13}
E=-(d_1+d_2)+\sqrt{d_1d_2}(x_i+x_i^{-1}),
\end{equation}
and $x_1x_2=1$. The equation  (\ref{boun12}) changes to
\begin{eqnarray}\label{boun14}
&&\left[E+D_1-D_1x_1\sqrt{d_1/d_2}\right]\left[(E+D_2)
-D_2x_1\sqrt{d_2/d_1}\right]x_1^{-2L+2}\cr &&
=\left[(E+D_1)x_1-D_1\sqrt{d_1/d_2}\right]\left[(E+D_2)
x_1-D_2\sqrt{d_2/d_1}\right].
\end{eqnarray}
Equation (\ref{boun14}) can be written as a polynomial equation of
order 2L, so it has $2L$ solutions. Two obvious solutions of the
(\ref{boun14}) are $x=\pm 1$. But, these generally do not
correspond to eigenvalues and eigenvectors. In fact for these
solutions, $x_1$ and $x_2$ are the same, so that (\ref{boun8})
should be modified to $C_k= (a +b k)(-1)^k$ and it is not
difficult to see that these do not fulfill the boundary conditions
unless $a=b=0$. It will be shown that two cases may occur, either
both solutions are phases then $|x_1|=|x_2|=1$, or both of them
are real . Except for the solutions $\pm 1$, one of the real
solutions is greater than one, which we take the be $x_1$. For the
phase solution, $x=\exp(i\theta)$
\begin{equation}\label{boun15}
E=-(d_1+d_2)+2\sqrt{d_1d_2}\cos \theta.
\end{equation}
Among these set of eigenvalues the biggest one is
$E=-(\sqrt{d_1}-\sqrt{d_2})^2$. So if there is no other solution
except for the phase solutions, the relaxation time is
\begin{equation}\label{boun16}
\tau_0=\frac{1}{ (\sqrt{d_1}-\sqrt{d_2})^2}.
\end{equation}
Now, let's search for the real solutions, if those exist. Of
course, by real solutions it is meant real solutions besides the
trivial solutions $\pm 1$.  If $x$ is a solution to
(\ref{boun14}), $x^{-1}$ is another solution to it. So it is
sufficient to seek the solutions with $|x|>1$. In the
thermodynamic limit, and for $|x|>1$, (\ref{boun14}) is
simplified to
\begin{equation}\label{boun16-2}
\left[(E+D_2)
x-D_2\sqrt{d_2/d_1}\right]\left[(E+D_1)x-D_1\sqrt{d_1/d_2}\right]=0
\end{equation}
If such a solution exists, then
\begin{equation}\label{boun17}
E=-(d_1+d_2)+\sqrt{d_1d_2}(x+x^{-1}).
\end{equation}
for any positive $x$, $E$ obtained from (\ref{boun17}) is greater
than the eigenvalue obtained from (\ref{boun15}), and so the
system relaxes to its stationary state slower. If all of the
solutions for (\ref{boun14}) are phases we call the system is in
fast phase, and if there is real solution, the system is in the
slow phase. Depending on parameters, there may be more than one
real solutions, and the system may be in the slow or slower phase.
The solutions of first bracket in (\ref{boun16-2}) is
$x=\sqrt{d_2/d_1}$, and $(d_1-D_2)/\sqrt{d_1d_2}$. The first
solution gives $E=0$, which is related to stationary state. We
obtained these solutions assuming $|x|>1$. So,
$x=(d_1-D_2)/\sqrt{d_1d_2}$ is a solution provided that
\begin{equation}\label{boun18}
{\cal A}:=\frac{d_1-D_2}{\sqrt{d_1d_2}}>1
\end{equation}
Similarly there may be a solution for the second bracket in
(\ref{boun16-2}), $x= (d_2-D_1)/\sqrt{d_1d_2}$, provided that
\begin{equation}\label{boun19}
{\cal B}:=\frac{d_2-D_1}{\sqrt{d_1d_2}}>1
\end{equation}
 The above results are summarized in the
figure 1. Depending on the magnitudes of ${\cal A}$, ${\cal B}$,
there are five regions.  If both ${\cal A}$, ${\cal B}$ are less
than one, then the relaxation time is given by
(\ref{boun16})(region $I$). Changing continuously the reaction
rates at the bulk or at the boundaries or the densities, the
relaxation time may changes discontinuously. This is the dynamical
phase transition. If one of ${\cal A}$, and ${\cal B}$ become
greater than one, then the relaxation time is given by $\tau_{
\cal A}$ or $\tau_{ \cal B}$
\begin{eqnarray}\label{boun20}
&&\tau_{ \cal A}:=\frac{d_1-D_2}{D_2(d_1-D_2-d_2)} \cr && \tau_{
\cal B}:=\frac{d_2-D_1}{D_1(d_2-D_1-d_1)}.
\end{eqnarray}
If ${\cal A}<1$, and ${\cal B}>1$ the system is in the phase $II$.
If ${\cal A}>1$, and ${\cal B}<1$ the system is in the phase
$III$. And finally if it is possible that  both ${\cal A}$, and
${\cal B}$ become greater than one, the relaxation time will be
given by ${\rm max}(\tau_{ \cal A}, \tau_{ \cal B})$. In the phase
$IV$, the relaxation time is $\tau_{ \cal B}$, and in the phase
$V$, the relaxation time is $\tau_{ \cal A}$.

Let's study  each of the three models ASEP, BCRW, and AKGP
separately.

\begin{itemize}
    \item 1. {\underline{\rm ASEP}}- To study the phase structure of this model,
    we should obtain ${\cal A}$, and ${\cal B}$.

    Let's assume $\omega_{32}<\omega_{23}$.
    Then $\omega_{32}<\sqrt{\omega_{23}\omega_{32}}$, and obviously
$$ \omega_{32}- \frac{\delta}{\rho_2}<\sqrt{\omega_{23}\omega_{32}}.$$
Now using the second equation of (\ref{21-2}), one arrives at
$$ d_1 -D_2<\sqrt{\omega_{23}\omega_{32}}= \sqrt{d_1d_2},\quad\Rightarrow \quad {\cal A}<1.$$
Now, let's assume $\omega_{23}<\omega_{32}$. Using (\ref{21}), and
the fact that $\beta $ is the extraction rate, and should be
positive, one gets
$$ \omega_{32}- \frac{\delta}{\rho_2}<\omega_{23}.$$
But $\omega_{23}< \sqrt{\omega_{23}\omega_{32}}$, so
$$ d_1 -D_2<\sqrt{\omega_{23}\omega_{32}}= \sqrt{d_1d_2},\quad\Rightarrow \quad {\cal A}<1.$$
With the similar reasoning, one gets ${\cal B}<1$.  Therefore the
only region in the space of parameters ${\cal A}$, and ${\cal B}$
available for the ASEP is the  region $I$. So, there is no
dynamical phase transition for ASEP, and the relaxation time is
given by
\begin{equation}\label{Aseprel}
\tau=\frac{\rho_1\rho_2}{(\rho_1\sqrt{\omega_{23}}-\rho_2\sqrt{\omega_{32}})^2}.
\end{equation}

    \item 2. {\underline{\rm BCRW}}- Let's first define
\begin{equation}\label{bcrw1}
\Omega :=\omega_{32}+ \frac{\rho_1}{1-\rho_1}\omega_{34}.
\end{equation}
     Using (\ref{21-3-2}), and
the fact that $\gamma $ is the extraction rate, and should be
positive, one gets
\begin{equation}\label{bcrw2}
\frac{\alpha(1-\rho_1)}{\rho_1}>\omega_{23}-(1-\rho_1)\Omega.
\end{equation}
Three cases may occur.
\begin{itemize}
\item  $\omega_{23}<\Omega $, and
$\alpha(1-\rho_1)/\rho_1<\omega_{23}-(1-\rho_1)\sqrt{\omega_{23}\Omega}$.
Then ${\cal A}<1$, and ${\cal B}>1$. Therefore,  the system is in
the phase $II$, and the relaxation time is given by $\tau_{\cal
B}$.

\item  $\omega_{23}>\Omega $, and
$ \beta <\omega_{23}- (1-\rho_1)\Omega $. Then ${\cal A}>1$, and
${\cal B}<1$. Therefore, the system is in the phase $III$, and the
relaxation time is given by $\tau_{\cal A}$.

\item  Otherwise the system is in the phase $I$, and the relaxation time is given by $\tau_0$,
(\ref{boun16}).

The regions with ${\cal A}>1$, and ${\cal B}>1$ is not available
for the BCRW. In summary, for the model BCRW, the parameter space
is restricted to the regions $I$, $II$, and $III$. So, there are
three distinct phases available for the system, and this model may
experience dynamical phase transitions.

\end{itemize}

    \item 3. {\underline {\rm AKGP}}- For this model we have
    \begin{eqnarray}\label{akgp1}
&&{\cal A}:=\sqrt{\frac{\omega_{13}}{\omega_{43}}}
-\frac{\beta}{\omega_{13}\omega_{43}}\cr &&  {\cal
B}:=\sqrt{\frac{\omega_{43}}{\omega_{13}}}
-\frac{\alpha}{\omega_{13}\omega_{43}}.
\end{eqnarray}
If ${\omega_{13}}>{\omega_{43}}$, Then ${\cal B}<1$. Depending on
the rates ${\omega_{13}}$ and ${\omega_{43}}$, and $\beta$, the
system may be in the phases $I$, or $III$. And if
${\omega_{13}}<{\omega_{43}}$, Then ${\cal A}<1$. Depending on the
rates ${\omega_{13}}$ and ${\omega_{43}}$, and $\alpha$, the
system may be in the phases $I$, or $II$. For the model AKGP, the
parameter space is restricted to the regions $I$, $II$, and $III$.
So, there are three distinct phases available for the system, and
the model  AKGP  may experience dynamical phase transitions.

\end{itemize}

\section{Double shock}

We define the state of a double shock on a one-dimensional lattice
with $L$ sites as
\begin{equation}\label{d1}
  |e_{m,k}\rangle= u^{\otimes m}\otimes v^{\otimes k}\otimes w^{\otimes
  (L-k-m)},\quad m+k\leq L
\end{equation}
where
\begin{equation}\label{d2}
 u:=\begin{pmatrix}1-\rho_1\cr \rho_1\end{pmatrix} \quad v:=\begin{pmatrix}1-\rho_2\cr
 \rho_2\end{pmatrix}\quad w:=\begin{pmatrix}1-\rho_3\cr
 \rho_3\end{pmatrix}.
\end{equation}
$|e_{m,k}\rangle$ represents a state for which the occupation
probability for the first $m$ sites is $\rho_1$, the occupation
probability for the next $k$ sites is $\rho_2$, and the occupation
probability for remaining sites is $\rho_3$. It should be noted
that this state represents a double shock, one shock at the site
$m$, and the other one at the site $m+k$. We call $k$ the width of
double-shock. We also assume that the three densities $\rho_1$,
$\rho_2$, and $\rho_3$ are different. The evolution of the shocks
in ASEP starting from a measure with several shocks and extra
particles at the shock positions is studied in \cite{BS}. In that
model, the measure is
\begin{equation}\label{d01}
  |e_{m,k}\rangle= u^{\otimes m}\otimes Z\otimes  v^{\otimes k}
  \otimes Z\otimes w^{\otimes
  (L-k-m)},\quad m+k\leq L,
\end{equation}
where
\begin{equation}\label{d02}
 Z:=\begin{pmatrix}0\cr 1\end{pmatrix}.
\end{equation}
It is shown that with a special choice of densities, time
evolution equation of this model is similar to that of some random
walkers on a lattice.

Although the states $|e_{m,k}\rangle,\quad (m,k=0,1,\cdots)$ do
not construct a complete basis, they are linearly independent
provided that $k$ the width of double-shock  can be zero only when
$m=0$ or $m=L$. This means that there should be at least one site
with the occupation probability $\rho_2$ between the sites with
the occupation probability $\rho_1$, and the sites with the
occupation probability $\rho_3$. For the single-shock linear
independency of $|e_{k}\rangle$s are obvious, but here $u$, $v$,
and $w$ are not linearly independent, So it is not obvious that
$|e_{m,k}\rangle$'s are linearly independent. To prove
$|e_{m,k}\rangle$'s are linearly independent, one must show that
if
\begin{equation}\label{d3}
  \sum_{k=0}^{L}\sum_{m=0}^{L-k}a^{mk}|e_{m,k}\rangle=0,
\end{equation}
then all $a^{mk}$'s must be zero. Let's define
\begin{equation}\label{d4}
 \tilde u:=\begin{pmatrix}\rho_1& \rho_1-1\end{pmatrix}, \quad  \tilde
 v:=\begin{pmatrix}\rho_2&
 \rho_2-1\end{pmatrix},\quad  \tilde w:=\begin{pmatrix}\rho_3&
 \rho_3-1\end{pmatrix}.
\end{equation}
Multiplying equation (\ref{d3}) from left hand side by
$|s\rangle^{\otimes L}$, one arrives at
\begin{equation}\label{d5}
  {\sum_{k=0}^{L}}^\prime\sum_{m=0}^{L-k}a^{mk}=0.
\end{equation}
Prime on the first summation denotes that $k=0$, only when $m=0$,
or $m=L$. Now, multiplying equation (\ref{d3}) from left hand
side by $\langle s |^{\otimes L-1}\otimes \tilde u$, and $\tilde
w\otimes \langle  s|^{\otimes L-1}$ one arrives at
\begin{eqnarray}\label{d6}
  &&\sum_{k=0}^{L}\sum_{m=0}^{L-k}a^{mk}- a^{L0}=0\cr
  &&\sum_{k=0}^{L}\sum_{m=0}^{L-k}a^{mk}- a^{00}=0.
\end{eqnarray}
(\ref{d5}) together with (\ref{d6}) gives $a^{L0}=a^{00}=0$. Now,
define $f^{r,n}$ through
\begin{equation}\label{d7}
  f^{rn}:=\tilde v^{\otimes r}\otimes s^{\otimes n}\otimes \tilde v^{\otimes
  (L-r-n)},
\end{equation}
where $r$, and $n$ are non-negative numbers and $n+r\leq L$.
Multiplying equation (\ref{d3}) from left hand side by $f^{r,1}$
and using $f^{r,1}|e_{m,k}\rangle= \delta^r_m \delta^1_k$ (for
$k\ne 0 $), one arrives at $a^{r1}=0$. Similarly, it can be shown
that all the coefficients $a^{mk}$ in (\ref{d3}) should be zero.
So, the set $|e_{m,k}\rangle$'s are linearly independent.

It was previously mentioned that there are three families of
stochastic one-dimensional non-equilibrium lattice models,
(ASEP,BCRW,AKGP), for which if the initial state is a linear
superposition of single-shock measures, at later times the state
of the system ${|\mathbb P\rangle }$  remains a linear combination
of shock measures. Among these, ASEP is the only model for which
double-shocks forms an invariant subspace, which means that if the
initial state is a linear superposition of the double-shock
measures, at later times the state of the system ${|\mathbb
P\rangle }$ remains a linear combination of double-shocks. It can
be shown that for the case of ASEP and on a one-dimensional
lattice with infinite sites,
\begin{eqnarray}\label{d8}
 {\mathcal H} |e_{m,k} \rangle&=& d_1 |e_{m-1,k+1} \rangle+d_2 |e_{m+1,k-1}\rangle+d_3 |e_{m,k-1} \rangle+
 d_4 |e_{m,k+1}\rangle\cr &&-(d_1+d_2+d_3+d_4) |e_{m,k}\rangle,\qquad k\geq 2
 \cr
 &&\cr
 {\mathcal H} |e_{m,1}\rangle&=& d_1 |e_{m-1,2}\rangle+d_4 |e_{m,2}\rangle-(d_1+d_4)
 |e_{m,1}\rangle,
\end{eqnarray}
where
\begin{equation}\label{d8-1}
  \frac{\omega_{23}}{\omega_{32}}=
  \frac{\rho_2(1-\rho_1)}{\rho_1(1-\rho_2)}=\frac{\rho_3(1-\rho_2)}{\rho_2(1-\rho_3)},
\end{equation}
and
\begin{eqnarray}\label{d8-2}
d_1= (\omega_{23}-
\omega_{32})\frac{\rho_1(1-\rho_1)}{\rho_2-\rho_1}\cr d_2=
(\omega_{23}-
\omega_{32})\frac{\rho_2(1-\rho_2)}{\rho_2-\rho_1}\cr d_3=
(\omega_{23}-
\omega_{32})\frac{\rho_2(1-\rho_2)}{\rho_3-\rho_2}\cr d_4=
(\omega_{23}- \omega_{32})\frac{\rho_3(1-\rho_3)}{\rho_3-\rho_2}
\end{eqnarray}

If the initial state is a linear combination of double-shocks,
then
\begin{equation}\label{d9}
{|\mathbb P\rangle}(t)=
\sum_{m=-\infty}^{\infty}\sum_{k=1}^{\infty}p_{mk}(t)
{|e_{mk}\rangle },
\end{equation}
where $p_{m,k}$ is the contribution of  the double-shock, $mk$ in
the state of the system.  Using (\ref{7}),(\ref{d9}), and also the
linear independency of $|e_{m,k}\rangle$'s, one can obtain the
evolution equation for $p_{mk}$'s.
\begin{eqnarray}\label{d10}
\dot p_{m,k}&=& d_1 p_{m+1,k-1}+d_2p_{m-1,k+1}+d_3 p_{m,k+1}+ d_4
p_{m,k-1}\cr &&-(d_1+d_2+d_3+d_4) p_{m,k},\qquad k\geq 2\cr
 &&\cr
 \dot p_{m,1}&=& d_2p_{m-1,2}+d_3 p_{m,2}-(d_1+d_4) p_{m,1}
\end{eqnarray}
Let's define
\begin{equation}\label{d11}
    q_k:=\sum_{m=-\infty}^{\infty}p_{mk}.
\end{equation}
Then $q_k$ is the contribution of all double-shocks with the
width, the distance between two shocks, $k$ in the state of the
system. Conservation of the probability, (\ref{5}), leads to
\begin{equation}\label{d11-1}
\sum_{k=1}^{\infty}q_{k}=1.
\end{equation}
The evolution equations for $q_k$'s are
\begin{eqnarray}\label{d12}
\dot q_{k}&=& D_1 q_{k-1}+D_2q_{k+1}-(D_1+D_2) q_{k},\qquad k\geq
2\cr
 &&\cr
 \dot q_{1}&=& D_2q_{2}-D_1 q_{1},
\end{eqnarray}
where
\begin{equation}\label{d13}
D_1:= d_1+d_4,\qquad D_2:= d_2+d_3.
\end{equation}
$D_1$, and $D_2$ can be written in terms of $\rho_1$ and $\rho_2$,
and the diffusion rates.

\begin{eqnarray}\label{d14}
D_1&=& (\omega_{32}- \omega_{23})
\{\frac{\rho_1(1-\rho_1)}{\rho_1-\rho_2}+
\frac{\rho_3(1-\rho_3)}{\rho_2-\rho_3}\}\cr D_2&=& (\omega_{32}-
\omega_{23}) \{ \frac{\rho_2(1-\rho_2)}{\rho_1-\rho_2}+
\frac{\rho_2(1-\rho_2)}{\rho_2-\rho_3}\}
\end{eqnarray}
Subtracting these
\begin{equation}\label{d14-1}
    D_2-D_1=
    \frac{\rho_2(1-\rho_2)(1+F)(1-F)^2}{\omega_{23}[(1-\rho_2)+F\rho_2][F(1-\rho_2)+\rho_2]}
\end{equation}
where $F:= \omega_{32}/\omega_{23}$, it is seen that $D_1<D_2$.
One  can easily obtain the steady state solution for $q_k$'s.
\begin{eqnarray}\label{d15}
    q_k&=& \left( \frac{D_1}{D_2}\right)^{k-1}q_1\cr
    q_1&=& 1- \frac{D_1}{D_2}.
\end{eqnarray}
To obtain the second equation of  (\ref{d15}), we have used
(\ref{d11-1}). Using  the fact that $D_1<D_2$, it is seen that
$q_k$ goes to zero for large $k$'s. This means that at the
stationary state the contributions of the double-shocks with
larger width are less and the main contribution comes from thin
double-shocks.  This is reminiscent of double-shocks in Burgers
equation. In fact, in the hydrodynamic limit double shock is not
stable and converges to single shock. Here, we show that in the
microscopic level,  and in the thermodynamic limit, the stationary
value of the contribution of double shocks with larger width
become vanishingly small.

\section{acknowledgments}
The authors would like to thank M. Khorrami, and F.H. Jafarpour
for useful discussions and helpful comments. A. A. also would like
to acknowledges E. Fooladvand for fruitful  discussions. We would
also like to thank the referee  for his (her) useful comments.

\newpage

Figure 1- Exact phase diagram of the ASEP, BCRW, and AKGP, in the
${\cal A}$, ${\cal B}$ plane. For the ASEP, the parameter space is
restricted to the region  $I$. For the models BCRW, and AKGP, the
parameter space is restricted to the regions $I$, $II$, and $III$.

Figure 1- Exact phase diagram of the ASEP, BCRW, and AKGP, in the
${\cal A}$, ${\cal B}$ plane. For the ASEP, the parameter space is
restricted to the region  $I$. For the models BCRW, and AKGP, the
parameter space is restricted to the regions $I$, $II$, and $III$.

\newpage
 \setlength{\unitlength}{1mm}
\begin{picture}(80,80)
\includegraphics{shockphase.eps} \put(21,20.6){$I$}
\put(57.8,21){$II$}\put(57.5,43){$IV$}\put(45.5,56.5){$V$}\put(19.3,52){$III$}\put(73,6){${\cal
B }$}\put(4.7,66){${\cal A }$}\put(34.8,5.3){${\bf 1
}$}\put(5.8,36){${\bf 1}$}
\end{picture}

\end{document}